\documentclass{SCIS2026}

\begin{document}
\ArticleType{RESEARCH PAPER}
\Year{2025}
\Month{January}
\Vol{68}
\No{1}
\DOI{}
\ArtNo{}
\ReceiveDate{}
\ReviseDate{}
\AcceptDate{}
\OnlineDate{}
\AuthorMark{}
\AuthorCitation{}
\title{Recursive Binary Identification with Differential Privacy and Data Tampering Attacks}{Recursive Binary Identification with Differential Privacy and Data Tampering Attacks}

\author[1,2]{Jimin WANG}{}
\author[3]{Jieming KE}{}
\author[1,2]{Jin GUO}{}
\author[4]{Yanlong ZHAO}{{ylzhao@amss.ac.cn}}


\address[1]{School of Automation and Electrical
Engineering, University of Science and Technology
Beijing, Beijing~100083}
\address[2]{Key Laboratory of Knowledge Automation for Industrial Processes, Ministry of Education, Beijing~100083}
\address[3]{Department of Information Engineering, University of Padova, Padova 35131, Italy}
\address[4]{State Key Laboratory of Mathematical Sciences, Academy of Mathematics and Systems Science, Chinese Academy of Sciences, Beijing~100190}

\abstract{In this paper, we consider the parameter estimation in a bandwidth-constrained sensor network communicating through an insecure medium. The
sensor performs a local quantization, and transmits a 1-bit message to an estimation center through a wireless medium where the transmission of information is vulnerable to attackers. Both eavesdroppers and data tampering attackers are considered in our setting. A differential privacy method is used to protect the sensitive information against eavesdroppers. Then, a recursive projection algorithm is proposed such that the estimation center achieves the almost sure convergence and mean-square convergence when quantized measurements, differential privacy, and data tampering attacks are considered in a uniform framework. A privacy analysis including the convergence rate with privacy or without privacy is given. Further, we extend the problem to multi-agent systems. For this case, a distributed recursive projection algorithm is  proposed with guaranteed almost sure and mean square convergence. A simulation example is provided to illustrate the effectiveness of the proposed algorithms.}


\maketitle

\section{Introduction}\label{sec:1}
As one of the most fundamental methods for data analysis, system identification has made a series of theoretical achievements, and has been ubiquitously employed in numerous fields, such as engineering systems, physical systems, social systems, biological systems, economic systems and many others. There are two challenges faced in  system identification problems. Quantization is one challenge faced in such problems. Due to cost constraints and limited bandwidth, sensors in many estimation problem often send quantized measurements rather than exact values. The system identification with quantized observations has drawn great research effort and experienced substantial advancement during the past decades. Compared with traditional system identification, quantized observations provide very limited information on system output signals, introduce nonlinearities that complicate system identification, and thus bring essential difficulties in system identification. Fundamental progress has been achieved in methodology development, identification algorithms, essential convergence properties, and applications \cite{Brockett2000,Wang2003, guo2013,marelli2013,zhao2018,Casini2016,Pouliquen2020,Jafari2012,WangY2021,WangY2022,WangY2026}.

Beyond quantization constraints, privacy and security are another challenges faced in estimation problems, and have attracted widespread attention in the past decades due to security concerns in practical applications \cite{Zhang2021,Wang2025}. For example, in many Internet of Things applications, sensors collect confidential information about the state of the dynamic system and send it to an authorized user, which can be a remote estimator/controller or a cloud server, through a wireless communication channel \cite{Cao2013}. However, due to the broadcast nature of the wireless medium, this confidential information may be leaked to eavesdroppers \cite{Aysal2008}. Many works have been done in privacy-preserving in estimation problems. For example, Leong et al. \cite{Leong2019} investigated transmission scheduling for remote state estimation through a packet-dropping channel in the presence of an eavesdropper. Tsiamis et al. \cite{Tsiamis2020} designed a coding scheme that used acknowledgment signals and applied linear time-varying transformations to the current and previous states. Yang et al. \cite{Yang2020} proposed an encoding mechanism that combines a linear transformation of measurements and artificial noises. Huang et al. \cite{Huang2021} studied an optimal encryption scheduling under energy constraints to ensure estimation accuracy in an infinite-time horizon. \cite{Shang2023} used linear encryption strategies to protect remote state estimation from eavesdropping for both stable and unstable systems. Zou et al. \cite{Zou2023} was concerned with the secure state estimation problem for a networked system with multirate measurements. Guo et al \cite{Guo2025} addressed the real-time state estimation problem for dynamic systems while protecting exogenous inputs by Cram$\acute{e}$r-Rao lower bound approach. However, existing relevant research mainly focus on the state estimation. These studies are often carried out when the system model is known, while it is likely to be unknown in practice. Therefore, it is necessary to identify the system model, which is the starting point of this paper and results obtained can provide support for the state estimation. Homomorphic encryption enables certain algebraic operations to be carried out on ciphertexts while ensuring the privacy of plaintexts, secure parameter identification of ARX systems with multi-participants has been studied based on this scheme in \cite{Tan2023}. However, this scheme usually involves modular exponentiation operations over massive integers, which result in heavy online computational overhead.  Differential privacy ensures that critical information, such as the gradient or state of an individual agent, cannot be inferred from the perturbed data even with access to arbitrary auxiliary information by carefully designing added noises. Because of its mathematical quantifiability and post-processing property, differential privacy has attracted widespread attention and has been studied in Kalman filtering \cite{Ny2014}, observer design \cite{McGlinchey2017}, distributed consensus \cite{WangJM2024}, distributed optimization \cite{WangYQ2024a,Wang2024a}, and distributed estimation \cite{Wang2023}. Recently, \cite{Nam2024} considered a private discrete distribution estimation problem with one-bit communication constraint. The privacy constraints are imposed with respect to the local differential privacy and the maximal leakage. The estimation error is quantified by the worst-case mean squared error. Besides privacy, data tampering attacks in the system identification with binary-valued observations is also important, and has been studied in \cite{GuoJ2021,GuoJ2025}.

Inspired by the above discussions, we consider both privacy protection and data tampering attacks in the system identification with binary-valued observations. The novel idea of this paper is that the estimation center estimates the parameter $\theta$ in almost sure and mean-square sense when privacy, data tampering attacks and quantization constraints are considered in a unified framework. The main contributions of the paper are as follows:
\begin{itemize}
\item This paper is the first to address the system identification under quantized measurements, privacy protection and data tampering attacks simultaneously. Regarding the communication costs, the sensor transmits only 1 bit of information to the remote estimation center at each time step compared to the existing works on the privacy preserving estimation problem \cite{Leong2019,Tsiamis2020,Yang2020,Huang2021,Shang2023,Zou2023,Guo2025}. Regarding the privacy protection, differential privacy method is used compared to \cite{guo2013,GuoJ2021,WangY2022}. Furthermore, data tampering attacks are considered compared to \cite{Nam2024}
\item A privacy preserving recursive projection algorithm with the time-varying thresholds is proposed such that the estimation center achieves the almost sure and mean-square convergence, which make it is possible to use the algorithm for the deterministic system under data tampering attacks compared to \cite{guo2013,WangY2021,WangY2026}. Besides, the effect of the differential privacy is shown by analyzing the convergence rate of the algorithm. The convergence rate reaches $O(\frac{1}{k})$, which is the best among those only considering quantization and data tampering attacks \cite{GuoJ2025}, and only considering privacy protection \cite{Wang2023}.
\item We further extend the problem to the multi-agent systems, and propose a privacy preserving distributed recursive projection algorithm such that each agent estimates the parameter $\theta$ in the almost sure and mean-square sense. Compared with \cite{WangY2021}, privacy protection and data tampering attacks are considered. Compared with \cite{Wang2023}, binary measurements and data tampering attacks  are considered.
\end{itemize}

The remaining parts of this paper are organized as follows. Section \ref{sec:2} describes the problem formulation, including differential privacy, binary-valued observations, and data tampering attacks. Section \ref{sec:3} gives the main results, including a recursive projection algorithm, convergence and privacy analysis, and a distributed recursive projection algorithm. One numerical simulation is performed in Section \ref{sec:5}.

\section{Problem Formulation}\label{sec:2}
\subsection{System model}
Consider the following system:
\begin{align}\label{system}
y_{k} = \varphi_{k}^{T}\theta,
\end{align}
where $\theta\in\Omega\subseteq\mathbb{R}^{p}$ is the unknown parameter to be estimated but time-invariant parameter vector of known dimension $p$, and $\varphi_k\in \mathbb{R}^{p}$ is the regressor vector composed of current and past input signals.

\subsubsection{Differential privacy}
In order to protect the sensitive information from the potential adversary, inspired by [34], we introduce the concepts about differential privacy.
\begin{definition}\label{adj} (Adjacency relationship) Given $\Delta_{k}>0$, two vectors $y_k$ and $y'_k$, $y_k$ and $y'_k$ are adjacent if
$\|y_k-y'_k\|\leq\Delta_k$.
\end{definition}
\begin{remark}
Definition \ref{adj} implies that two signal sets are adjacent if the measurement vector changes, which is a key component of the private implementation as it specifies which pieces of sensitive data must be made approximately indistinguishable to the potential adversary.
$\|y_k-y'_k\|$ is required to be bounded by an exponentially decaying function \cite{McGlinchey2017}, and constant in \cite{Wang2023}. However, in this work, we relax this constraint to enable privacy preservation across wider variations of adjacent datasets, which strengthens data preservation.
\end{remark}

\begin{definition}(Differential privacy) \cite{Ny2014} Given $\epsilon,\delta> 0$, a randomized mechanism $\mathcal{M}$ is $(\epsilon,\delta)$-differentially private if for any adjacent vectors $y_k$ and $y'_k$, and any set of outputs  $R\subset Range(\mathcal{M})$, such that
\begin{align}
\mathbb{P}(M(y_k) \in R)\leq e^\epsilon\mathbb{P}(M(y'_k) \in R)+\delta.
\end{align}
\end{definition}
\begin{remark}
The above inequality is standard in defining the differential privacy. Since it holds for any adjacent $y_k$ and $y'_k$, we can exchange $ M(y_k)$
with $M(y'_k)$,  and obtain $\mathbb{P}(M(y'_k) \in R)\leq e^\epsilon\mathbb{P}(M(y_k) \in R)+\delta$. Subtracting the two inequalities yields $1-e^\epsilon-\delta\leq(1-e^\epsilon)\mathbb{P}(M(y'_k) \in R)-\delta\leq\mathbb{P}(M(y'_k) \in R)-\mathbb{P}(M(y_k) \in R)\leq (e^\epsilon-1)\mathbb{P}(M(y_k) \in R)+\delta\leq e^\epsilon-1+\delta$, and hence $|\mathbb{P}(M(y'_k) \in R)-\mathbb{P}(M(y_k) \in R)|\leq e^\epsilon-1+\delta$. Since $e^\epsilon\approx 1 + \epsilon$ for small $\epsilon > 0$, it means that for sufficiently small $\epsilon$, $\delta>0$, the eavesdropper cannot distinguish $y_k$ from $y'_k$ based on the observation $M$. This demonstrates that privacy protection of $y_k$ is achieved.
\end{remark}
\subsubsection{Binary-valued observation}
Let $\omega_{k}$ be a privacy noise sequence. The private output $\tilde{y}_{k}=y_{k}+\omega_k$ is measured by a binary-valued sensor with the time-varying threshold of $\varphi_{k}^{T}\hat{\theta}_k$, where $\hat{\theta}_k$ is the estimate of $\theta$,  and is represented by the following indicator function:
\begin{equation}\label{sk}
s_k^0 = I_{\{\tilde{y}_k\leq \varphi_{k}^{T}\hat{\theta}_k\}}=\left\{\begin{array}{rl}
                            1,& \tilde{y}_k\leq \varphi_{k}^{T}\hat{\theta}_k;\\
                            0,& \mbox{otherwise}.
                          \end{array}\right.
\end{equation}
\begin{remark}
The time-varying threshold is used in (\ref{sk}) due to System (\ref{system}) is deterministic. If System (\ref{system}) is a stochastic one as shown in \cite{guo2013}, the threshold in (\ref{sk}) can be fixed, and the main results studied later still hold.  Here we take the deterministic system (\ref{system}) to show  the main results clearer.
\end{remark}
\subsubsection{Data tampering attacks}
$s_k^0$ is transmitted to the remote estimation center through a communication network, but subject to data tampering attacks. The outputs when attacker exists, denoted as $s_k$, are transmitted through an open wireless medium. The data tampering attacker model considered here is defined as
\begin{equation}\label{og}
   \left\{
   \begin{array}{lll}
     \mathbb{P}\{s_k=0|s_k^0=1\} =p; \\[.1cm]
     \mathbb{P}\{s_k=1|s_k^0=1\} =1-p; \\[.1cm]
     \mathbb{P}\{s_k=1|s_k^0=0\} =q; \\[.1cm]
     \mathbb{P}\{s_k=0|s_k^0=0\} =1-q,
   \end{array}
   \right.
\end{equation}
where $p,q\in[0,1]$.
\begin{figure}[H]
\centering
\includegraphics[width=8cm]{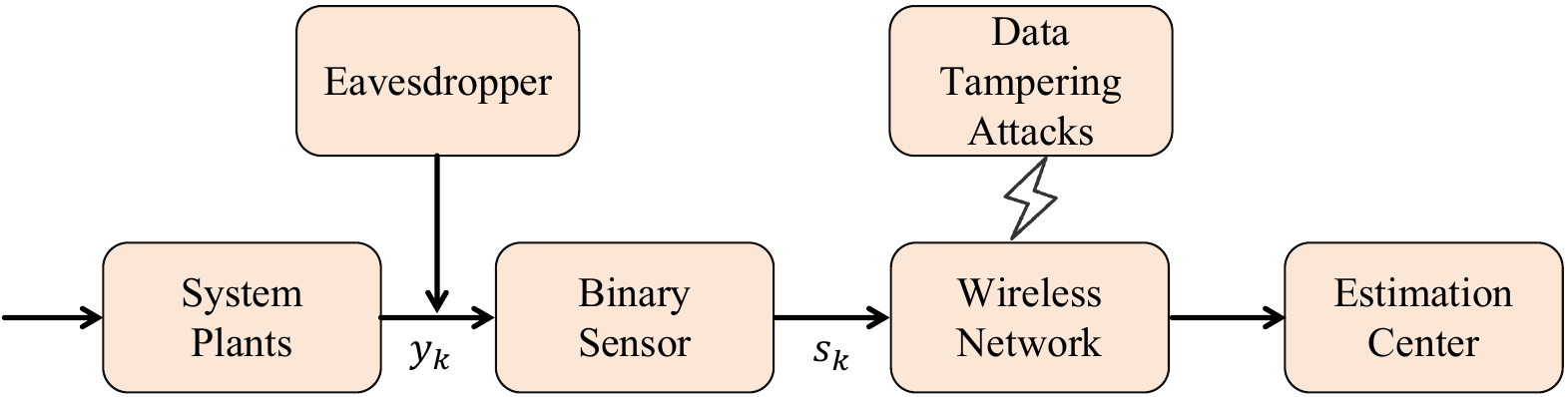}\\
\caption{System configuration}\label{fig-system}
\end{figure}
\subsection{Problem of interest}
In this paper, as shown in Figure \ref{fig-system},  we consider the differential privacy, data tampering attacks, and quantized measurements for parameter estimation in a unified framework, and aim to solve the following three problems.
\begin{itemize}
\item Under privacy protection, data tampering attacks and quantized measurements, we design an algorithm such that the estimation center estimates the parameter $\theta$.
\item Analyze the convergence rate of the algorithm, and show the effect of the differential privacy.
\item Extend to the multi-agent systems, and propose a privacy preserving distributed recursive projection algorithm such that each agent estimates the parameter $\theta$.
\end{itemize}
\section{Main Results}\label{sec:3}
\subsection{Algorithm}
\begin{definition}\label{DF1}
Let $\Omega  \subseteq \mathbb{R}^n$ be a convex compact set. The projection operator $\Pi_{\Omega }(\cdot)$ is defined by
$\Pi_{\Omega }(x) = \arg\min_{\omega \in \Omega } \| x - \omega \|, \quad \forall x \in \mathbb{R}^n.$
\end{definition}

\begin{proposition}
The projection operator given by Definition \ref{DF1} follows
\begin{align}\label{pr1}
\|\Pi_{\Omega }(x)-\Pi_{\Omega }(y)\|\leq\|x-y\|, x, y \in \mathbb{R}^{n}.
\end{align}
\end{proposition}

\begin{assumption}\label{A2}
The unknown parameter satisfies $\theta \in \Omega \subseteq \mathbb{R}^n$, where $\Omega$ is a convex compact set. And denote $\eta=\sup_{w\in\Omega}\|w\|$.
\end{assumption}

Based on the above analysis, we propose a privacy preserving recursive projection algorithm for system identification with binary observations, see Algorithm 1.

\begin{algorithm}
\caption{A privacy preserving recursive projection algorithm}
\begin{algorithmic}
\STATE Initial the estimate $\hat{\theta}_1 \in \Omega$, the step-size sequence $\{b_k\}_{k \geq 1}$, a constant gain parameter $\beta > 0$

\STATE For $k = 1, 2, \ldots$
\STATE Compute
\begin{align}\label{10}
\tilde{s}_{k} &= \beta(1 - (p + q)) \left( (1 - (p + q))F(0) + q - s_{k} \right).
\end{align}

\STATE Update parameter estimate as follows
\begin{align}\label{11}
\hat{\theta}_{k+1} = \Pi_{\Omega} \left\{ \hat{\theta}_k + b_k \varphi_k \tilde{s}_{k} \right\}.
\end{align}
\end{algorithmic}
\end{algorithm}

\subsection{Convergence analysis}

In this section, we show the estimation center's estimation error. Define $F(\cdot)$ as the probability distribution function of the privacy noise, and the natural filtration $\mathcal{F}_{k}$ as
\begin{align}
\mathcal{F}_{k}=\sigma\{\varphi_{0}, \ldots, \varphi_{k+1}, \omega_{0}, \ldots, \omega_{k}\}, k\geq0.
\end{align}
Based on the law of total probability and the data tampering attacks method in (\ref{og}), the conditional probability of observing $s_{k} = 0$ can be computed as
\begin{align}\label{IA5}
&\mathbb{P}\left(s_{k}=0|\mathcal{F}_{k-1}\right)\cr
=&\mathbb{P}\left(s_{k}^0=0|\mathcal{F}_{k-1}\right)\cdot \mathbb{P}\left(s_{k}=0|s_{k}^0=0,\mathcal{F}_{k-1}\right)+\mathbb{P}\left(s_{k}^0=1|\mathcal{F}_{k-1}\right)\cdot \mathbb{P}\left(s_{k}=0|s_{k}^0=1,\mathcal{F}_{k-1}\right)\cr
=&pF(\varphi_{k}^{T}\hat{\theta}_k-\varphi_{k}^{T}\theta)+(1-q)[1-F(\varphi_{k}^{T}\hat{\theta}_k-\varphi_{k}^{T}\theta)]\cr
=&(p+q-1)F(\varphi_{k}^{T}\hat{\theta}_k-\varphi_{k}^{T}\theta)+1-q.
\end{align}
The conditional probability of $s_{k} = 1$ is given by
\begin{align}\label{IA6}
&\mathbb{P}\left(s_{k}=1|\mathcal{F}_{k-1}\right)\cr
=&1-\mathbb{P}\left(s_{k}=0|\mathcal{F}_{k-1}\right)\cr
=&(1-(p+q))F(\varphi_{k}^{T}\hat{\theta}_k-\varphi_{k}^{T}\theta)+q.
\end{align}
From (\ref{IA5})-(\ref{IA6}), it can be seen that when $p+q=1$, the dependence on the conditional distribution $F(\cdot)$ vanishes. In this case, the distribution of the observations becomes independent of $\theta$, making the parameter identification impossible. Therefore, a necessary condition of identifiability is: $p+q\neq1$. Then, at each time step $k$, the conditional expectation of the observed binary output $s_{k}$ is given by
\begin{align}\label{al8}
\mathbb{E}\left[s_{k}|\mathcal{F}_{k-1}\right]=(1-(p+q))F(\varphi_{k}^{T}\hat{\theta}_k-\varphi_{k}^{T}\theta)+q.
\end{align}

The following assumptions are also needed.

\begin{assumption}\label{A3}
(a) The regressor sequence $\{ \varphi_k \}$ satisfies
\begin{align}\label{12}
\sup_{k \geq 1} \| \varphi_k \| \leq M < \infty.
\end{align}
(b) There exist an integer $h \geq p$ and a constant $\delta > 0$ such that
\begin{align}\label{13}
\frac{1}{h} \mathbb{E} \left[ \sum_{l=k}^{k+h-1} \varphi_l \varphi_l^T \bigg| \mathcal{F}_{k-1} \right] \geq \delta_{\varphi} I, \quad \forall k \geq 1,
\end{align}
where $I$ is the $p \times p$ identity matrix.
\end{assumption}

\begin{assumption}\label{A4}
The noise $ \omega_{k} $ is independent of $\varphi_{0}, \ldots, \varphi_{k}, \omega_{0}, \ldots, \omega_{k-1}$, and the density function $f(\cdot)$ of the noise $\omega_k$ satisfies
\begin{align}
\underline{f} :=  \inf_{|x| \leq M\eta} f(x) > 0.
\end{align}
\end{assumption}

\begin{assumption}\label{A5}
The step-size sequence $\{ b_k \}$ is monotonically decreasing and satisfies
\begin{align}
\sum_{k=1}^{\infty} b_k = \infty, \quad \lim_{k \to \infty} b_k = 0, \quad \text{and} \quad b_k = O(b_{k+1}).
\end{align}
\end{assumption}

\begin{remark}
Assumption \ref{A3} places conditions on the input sequence $\{ \varphi_k \}$. Condition (\ref{13}), known as the ``conditionally expected sufficiently rich condition," ensures that the regressor sequence contains enough variability to allow parameter identifiability.
\end{remark}

\begin{remark}
Assumption \ref{A5} is standard and commonly used in stochastic approximation algorithms. These conditions ensure that the step size remains effective over time, decays gradually, and avoids premature vanishing. For example, if $b_k=\frac{1}{(k+1)^{\alpha}}, \alpha\in(0,1]$, then Assumption \ref{A4} holds.  The constant $\underline{f}$ in Assumption \ref{A4} provides a uniform lower bound, ensuring that the noise remains sufficiently dispersed in the estimation region.
\end{remark}

\begin{lemma}\cite{Polyak1987}\label{lemma1}
Let $\{p_k\}$, $\{q_k\}$ and $\{\alpha_k\}$ be real sequences satisfying $p_{k+1}\leq(1-q_k)p_k+\alpha_k$, where
$0<q_k\leq1$, $\sum_{k=1}^{\infty} q_k=\infty$, $\alpha_k\geq0$, and $\lim_{k\rightarrow\infty}\frac{\alpha_k}{q_k}=0$. Then, $\limsup_{k\rightarrow\infty}p_k\leq0$. In particular, if $p_k\geq0$, then $\lim_{k\rightarrow\infty}p_k=0$.
\end{lemma}

Denote the  estimation center's estimation error as $\tilde{\theta}_k = \hat{\theta}_k - \theta$, $k = 1, 2, \ldots$. Then, we have the following lemma.
\begin{lemma}\label{lemma2}
Given $l \in \mathbb{N}$, if Assumptions \ref{A2}, \ref{A4} hold, then
\begin{align*}
\|\tilde{\theta}_{k+l}-\tilde{\theta}_k\|=O\left(b_{k+l}\right).
\end{align*}
\end{lemma}
{\bf Proof.} Since $0\leq (1 - (p + q))F_k(\varphi_{k}^{T}\hat{\theta}_k-\varphi_k^T \theta) + q\leq1$, we have $|\tilde{s}_{k}|\leq\beta$.
This together with (\ref{pr1}) and the condition $\|\varphi_k\|\leq M$, it is obtained that
\begin{align}\label{lm1}
\|\hat{\theta}_{l+1}-\hat{\theta}_l\|\leq b_{l+1}\|\varphi_{l+1}\tilde{s}_{l+1}\|\leq b_{l+1}\beta M, \forall l\geq1.
\end{align}
Note that
\begin{align*}
\|\tilde{\theta}_{k+l}-\tilde{\theta}_k\|=&\|\hat{\theta}_{k+l}-\hat{\theta}_k\|
=\|\sum_{j=1}^{l}\left(\hat{\theta}_{k+j}-\hat{\theta}_{k+j-1}\right)\|\cr
\leq&\sum_{j=1}^{l}\|\hat{\theta}_{k+j}-\hat{\theta}_{k+j-1}\|.
\end{align*}
This together with (\ref{lm1}) and Assumption \ref{A5} implies the lemma.

The following theorem establishes the estimation center's estimation error converges to 0 in  the almost sure and mean-square sense.

\begin{theorem}\label{theorem}
Under Assumptions \ref{A2}-\ref{A5}, if $p+q\neq1$, then the parameter estimate generated by Algorithm 1 satisfies
\begin{align}
\lim_{k \to \infty} \mathbb{E} \| \tilde{\theta}_k \|^2 = 0.
\end{align}
Moreover, if $\sum_{k=1}^{\infty} b_k^2 < \infty$, then the estimate $\hat{\theta}_k$ also converges almost surely to the true parameter, i.e.,
\begin{align}
\lim_{k \to \infty} \tilde{\theta}_k = 0, \quad \text{a.s.}
\end{align}
\end{theorem}
{\bf Proof.} By $\tilde{s}_k^2 \leq \beta^2$, (\ref{pr1}), (\ref{10}) and Assumption \ref{A2}, we have
\begin{align}\label{th1}
\|\tilde{\theta}_{k+1}\|^2 &\leq \|\tilde{\theta}_k\|^2 + 2b_k \tilde{s}_{k} \varphi_k^T \tilde{\theta}_k + b_k^2 \|\varphi_k\|^2 \beta^2 \cr
&= \|\tilde{\theta}_k\|^2 + 2b_k \tilde{s}_{k} \varphi_k^T \tilde{\theta}_k + O(b_k^2).
\end{align}
From (\ref{al8}) and (\ref{10}), it follows that
\begin{align}\label{th2}
\mathbb{E}[\tilde{s}_{k} | \mathcal{F}_{k-1}]
= \beta(1 - p - q)^2 \left( F(0) - F(\varphi_{k}^{T}\hat{\theta}_k - \varphi_k^T \theta) \right).
\end{align}
This together with Assumption \ref{A4} and the differential mean value theorem, it leads to
\begin{align}\label{th3}
&\mathbb{E} \left[ 2b_k \tilde{s}_{k} \varphi_k^T \tilde{\theta}_k | \mathcal{F}_{k-1} \right] \cr
= &2b_k \varphi_k^T \tilde{\theta}_k \mathbb{E}[\tilde{s}_{k} | \mathcal{F}_{k-1}] \cr
=& 2b_k \varphi_k^T \tilde{\theta}_k \beta(1 - p - q)^2 \left( F(0) - F(\varphi_{k}^{T}\hat{\theta}_k - \varphi_k^T \theta) \right) \cr
=& -2b_k \beta(1 - p - q)^2 f_k(\xi_k) \tilde{\theta}_k^T \varphi_k \varphi_k^T \tilde{\theta}_k \cr
\leq& -2b_k \beta(1 - p - q)^2 \underline{f} \tilde{\theta}_k^T \varphi_k \varphi_k^T \tilde{\theta}_k,
\end{align}
where $\xi_k$ is in the interval between $0$ and $\varphi_{k}^{T}\hat{\theta}_k- \varphi_k^T \theta$ such that $F(0) - F(\varphi_{k}^{T}\hat{\theta}_k - \varphi_k^T \theta) = f_k(\xi_k) \tilde{\theta}_k^T \varphi_k \varphi_k^T \tilde{\theta}_k$. Taking the conditional expectation on both sides of (\ref{th1}) and substituting (\ref{th3}) into it, we can obtain
\begin{align}\label{th4}
&\mathbb{E}\left[ \|\tilde{\theta}_{k+1}\|^2| \mathcal{F}_{k-1} \right]  \cr
\leq&  \|\tilde{\theta}_k\|^2 + 2b_k \mathbb{E} \left[ 2b_k\tilde{s}_{k} \varphi_k^T \tilde{\theta}_k | \mathcal{F}_{k-1} \right]  + O(b_k^2) \cr
\leq &  \|\tilde{\theta}_k\|^2 -2b_k \beta(1- p-q)^2 \underline{f} \tilde{\theta}_k^T \varphi_k \varphi_k^T \tilde{\theta}_k + O(b_k^2).
\end{align}
Taking the expectation on both sides of (\ref{th4}) implies
\begin{align}\label{th4a}
\mathbb{E} \|\tilde{\theta}_{k+1}\|^2
\leq  \mathbb{E}\|\tilde{\theta}_k\|^2 -2b_k \beta(1- p-q)^2 \underline{f} \mathbb{E}\left[\tilde{\theta}_k^T \varphi_k \varphi_k^T \tilde{\theta}_k\right] + O(b_k^2).
\end{align}
By iterating (\ref{th4a}) $h$ times and noting $b_k = O(b_{k+1})$, we obtain
\begin{align}\label{th5}
&\mathbb{E} \|\tilde{\theta}_{k+h}\|^2 \cr
\leq& \mathbb{E} \|\tilde{\theta}_k\|^2 - 2\beta(1 - p - q)^2 \underline{f} \mathbb{E} \left[ \sum_{l=k}^{k+h-1} \left[ b_l \tilde{\theta}_l^T \varphi_l \varphi_l^T \tilde{\theta}_l \right] \right] + O(b_{k+h}^2) \cr
\leq& \mathbb{E} \|\tilde{\theta}_k\|^2 - 2\beta(1 - p - q)^2 \underline{f} \mathbb{E} \left[ \sum_{l=k}^{k+h-1} \left[ b_l \tilde{\theta}_k^T \varphi_l \varphi_l^T \tilde{\theta}_k \right] \right] \cr
&- 2\beta(1 \!-\! p \!-\! q)^2 \underline{f} \mathbb{E} \left[ \sum_{l=k}^{k+h-1} \left[ b_l (\tilde{\theta}_l \!-\! \tilde{\theta}_k)^T \varphi_l \varphi_l^T (\tilde{\theta}_l \!-\! \tilde{\theta}_k) \right] \right] + O(b_{k+h}^2).
\end{align}
By Lemma \ref{lemma2} and (\ref{12}), the last two terms of (\ref{th4}) are of order $O(b_{k+h}^2)$. In addition,
\begin{align}\label{th6}
\mathbb{E} \left[ \sum_{l=k}^{k+h-1} \left[ b_l \tilde{\theta}_k^T \varphi_l \varphi_l^T \tilde{\theta}_k \right] \right] &= \mathbb{E} \left[ \tilde{\theta}_k^T \mathbb{E} \left[ \sum_{l=k}^{k+h-1} b_l \varphi_l \varphi_l^T \bigg| \mathcal{F}_{k-1} \right] \tilde{\theta}_k \right].
\end{align}
By Assumption \ref{A3}, we have
\begin{align}\label{th7}
&\mathbb{E} \left[ \sum_{l=k}^{k+h-1} b_l \varphi_l \varphi_l^T \bigg| \mathcal{F}_{k-1} \right] \cr
\geq & b_{k+h-1} \mathbb{E} \left[ \sum_{l=k}^{k+h-1} \varphi_l \varphi_l^T \bigg| \mathcal{F}_{k-1} \right]\cr
\geq &\delta_{\varphi} h b_{k+h-1}I.
\end{align}
Substituting (\ref{th6}) and (\ref{th7}) into (\ref{th5}) yields
\begin{align}\label{th8}
&\mathbb{E} \|\tilde{\theta}_{k+h}\|^2\cr
\leq& \mathbb{E} \|\tilde{\theta}_k\|^2 - 2\beta(1 - p - q)^2 \underline{f} \delta_{\varphi} h b_{k+h-1} \mathbb{E} \|\tilde{\theta}_k\|^2 + O(b_{k+h}^2) \cr
=& \left( 1 - 2\beta(1 - p - q)^2 \underline{f} \delta_{\varphi} h b_{k+h-1} \right) \mathbb{E} \|\tilde{\theta}_k\|^2 + O(b_{k+h}^2).
\end{align}
Then, based on Lemma \ref{lemma1} and Assumption \ref{A5}, and noting $\sum_{k=1}^{\infty} b_k = \infty$ and $\lim_{k \to \infty} b_k= 0$, it follows that $\lim_{k \to \infty} \mathbb{E} \|\tilde{\theta}_k\|^2 = 0$.

On the other hand, by (\ref{th4}) we have $\mathbb{E}[\|\tilde{\theta}_{k+1}\|^2 | \mathcal{F}_{k-1}] \leq \|\tilde{\theta}_k\|^2 + O(b_k^2)$, which together with Lemma 1.2.2 in \cite{chen2002} and $\sum_{k=1}^{\infty} b_k^2 < \infty$ implies that $\|\tilde{\theta}_k\|$ converges to a bounded limit a.s. Note that $\lim_{k \to \infty} \mathbb{E}\|\tilde{\theta}_k\|^2 = 0$. Then,  by Theorem 5.2.1 of \cite{Gut2005}, $\tilde{\theta}_k$ almost surely converges to $0$. \hfill$\square$
\begin{remark}
Theorem \ref{theorem} shows that the estimation center's estimation error converges to 0 in mean-square and almost sure sense when 1 bit of information transmissions, privacy protection and data tampering attacks are considered simultaneously.
\end{remark}

\subsection{Privacy analysis}
Let $\omega_{k}$ be a Gaussian noise sequence with zero mean, and variance $\sigma^{2}$. Based on the post-processing property of differential privacy (see \cite{Ny2014,Wang2023}), we have the following theorem.
\begin{theorem}
If the privacy noise satisfies $\sigma=\frac{\Delta_k}{2\epsilon}(\mathcal{Q}^{-1}(\delta)+\sqrt{(\mathcal{Q}^{-1}(\delta))^{2}+2\epsilon})$ with $\mathcal{Q}(\delta):=\frac{1}{\sqrt{2\pi}}\int_{\delta}^{\infty}e^{-\frac{u^2}{2}}{\rm d}u$, then each iteration of Algorithm 1 is $
(\epsilon,\delta)$-differentially private.
\end{theorem}
{\bf Proof.} By Theorem 3 and post-processing property (Theorem 1) in \cite{Ny2014}, each iteration of Algorithm 1 is $
(\epsilon,\delta)$-differentially private.   \hfill$\square$
\begin{remark}
 If $\Delta_k$ is constant, and the privacy noises follow Laplace mechanism, then the privacy results reduce to the ones in \cite{Wang2023}. Different from  \cite{Wang2023}, we relax the adjacency relationship, and further enhance the privacy level $\epsilon$. Especially under appropriate conditions on $\Delta_k$, infinite iterations of privacy protection can be achieved.
\end{remark}
\subsection{The effect of the privacy on the convergence rate}
In this subsection, we show the effect of the privacy on the convergence rate by establishing the convergence rate of Algorithm 1 with privacy or without privacy. To do this, we take $b_k=\frac{1}{k}$ as an example, the similar analysis can be reached for  $b_k=\frac{1}{k^{b}}, 0<b<1$.
\begin{lemma}\cite{Wang2024a}\label{lemma3}
For $0<b\leq1$, $a>0$, $k_0\geq0$ and sufficiently large $l$, we have
\begin{align}
\begin{array}{ll}
\prod_{i=l}^{k}\left(
1-\frac{a}{(i+k_0)^b}\right)\leq
\left\{
\begin{array}{ll}
\left(\frac{l+k_0}{k+k_0}\right)^{a},&b=1;\cr
e^{\frac{a}{1-b}((l+k_0)^{1-b}-(k+k_0+1)^{1-b})},&b\in(0,1).
\end{array}
\right.
\end{array}
\end{align}
\end{lemma}
\begin{lemma}\cite{Wang2024a}\label{lemma4}
For the sequence $h_k$, assume that (i) $h_k$ is positive and monotonically increasing; (ii) $\ln h_k = o(\ln k)$. Then, for
real numbers $a_1$, $a_2$, $\chi$, and any positive integer $p$, we have
\begin{align*}
& \sum_{l=1}^{k} \prod_{i=l+1}^{k}\left(1-\frac{a_1}{i+a_2}\right)^{p} \frac{h_l}{l^{1+\chi}} =
\begin{cases}
O\left(\frac{1}{k^{pa_1}}\right), & pa_1 < \chi, \\
O\left(\frac{h_k\ln k}{k^{\chi}}\right), & pa_1 = \chi,\\
O\left(\frac{h_k}{k^{\chi}}\right), & pa_1 > \chi.
\end{cases} &
\end{align*}
\end{lemma}
\subsubsection{Convergence rate with privacy}
\begin{theorem}\label{privacy}
 Under Assumptions \ref{A2}-\ref{A4}, if the gain parameter satisfies $\beta>\frac{1}{2(1-p-q)^{2}\underline{f}\delta_{\varphi}}$, then the mean-square convergence rate of Algorithm 1 is given as follows.
 \begin{align}
\mathbb{E} \| \tilde{\theta}_k \|^2 = O\left(\frac{1}{k}\right).
\end{align}
\end{theorem}
{\bf Proof.} Setting $\alpha=2\beta(1-p-q)^{2}\underline{f}\delta_{\varphi}$. Then, from (\ref{th8}) it follows that
\begin{align}
\mathbb{E} \| \tilde{\theta}_k \|^2 \leq&\left( 1 - \alpha \sum_{l=k-h}^{k-1} \frac{1}{l} \right) \mathbb{E} \|\tilde{\theta}_{k-h}\|^2 + O(\frac{1}{k^2})\cr
\leq&\left( 1 - \frac{\alpha h}{k} \right) \mathbb{E} \|\tilde{\theta}_{k-h}\|^2 + O(\frac{1}{k^2})\cr
\leq&\prod_{l=0}^{\lfloor\frac{k}{h}\rfloor-1}\left(1-\frac{\alpha h}{k-lh}\right) \mathbb{E} \|\tilde{\theta}_{k-\lfloor\frac{k}{h}\rfloor h}\|^2+
\sum_{l=0}^{\lfloor\frac{k}{h}\rfloor-1}\prod_{m=0}^{l-1}\left(1-\frac{\alpha h}{k-mh}\right) O\left(\frac{1}{(k-lh)^{2}}\right)\cr
\leq&\prod_{l=\kappa+1}^{\lceil\frac{k}{h}\rceil}\left(1-\frac{\alpha h}{lh}\right) \mathbb{E} \|\tilde{\theta}_{k-\lfloor\frac{k}{h}\rfloor h}\|^2+
\sum_{l=1}^{\lfloor\frac{k}{h}\rfloor}\prod_{m=\kappa+l+1}^{\lfloor\frac{k}{h}\rfloor}\left(1-\frac{\alpha h}{mh}\right) O\left(\frac{1}{(lh)^{2}}\right)\cr
\leq&\prod_{l=\kappa+1}^{\lceil\frac{k}{h}\rceil}\left(1-\frac{\alpha}{l}\right) \mathbb{E} \|\tilde{\theta}_{k-\lfloor\frac{k}{h}\rfloor h}\|^2+
\sum_{l=1}^{\lfloor\frac{k}{h}\rfloor}\prod_{m=\kappa+l+1}^{\lfloor\frac{k}{h}\rfloor}\left(1-\frac{\alpha }{m}\right) O\left(\frac{1}{l^{2}}\right),
\end{align}
where $\kappa=\lceil\frac{k}{h}\rceil-\lfloor\frac{k}{h}\rfloor$. Note that $\beta>\frac{1}{2(1-p-q)^{2}\underline{f}\delta_{\varphi}}$. Then, $\alpha>1$, and thus by Lemmas \ref{lemma3} and \ref{lemma4}, we have $\mathbb{E} \| \tilde{\theta}_k \|^2 = O\left(\frac{1}{k}\right).$
\subsubsection{Convergence rate without privacy}
When privacy is not considered, (\ref{10}) is replaced  by
\begin{align}\label{wp1}
\tilde{s}_{k} &= \beta(1 - (p + q))\left( q- s_{k} \right),
\end{align}
where
\begin{align}\label{wp2}
\mathbb{E}\left[s_{k}\right]=(1-(p+q))I_{\{\varphi_{k}^{T}\hat{\theta}_k-\varphi_{k}^{T}\theta\geq0\}}+q,
\end{align}
and (\ref{wp2}) follows from the same reason of (\ref{al8}).
\begin{theorem}\label{noprivacy}
 Under Assumptions \ref{A2}-\ref{A3}, if the gain parameter satisfies $\beta>\frac{\eta M}{ 2\delta_{\varphi} (1 - p - q)^2}$, then the mean-square convergence rate of the algorithm without privacy ((\ref{11}) and (\ref{wp1})) is given as follows.
 \begin{align}
\mathbb{E} \| \tilde{\theta}_k \|^2  = O\left(\frac{1}{k}\right).
\end{align}
\end{theorem}
{\bf Proof.} By $\tilde{s}_k^2 \leq \beta^2$ and (\ref{th1}), we have
\begin{align}\label{wp3}
\|\tilde{\theta}_{k+1}\|^2 &\leq \|\tilde{\theta}_k\|^2 + \frac{2}{k}\tilde{s}_{k} \varphi_k^T \tilde{\theta}_k + b_k^2 \|\varphi_k\|^2 \beta^2 \cr
&= \|\tilde{\theta}_k\|^2 + \frac{2}{k} \tilde{s}_{k} \varphi_k^T \tilde{\theta}_k + O(\frac{1}{k^2}).
\end{align}
From (\ref{wp1}) and (\ref{wp2}), it follows that
\begin{align}\label{wp4}
\mathbb{E}[\tilde{s}_{k}] = -\beta(1 - p - q)^2 I_{\{\varphi_{k}^{T}\hat{\theta}_k-\varphi_{k}^{T}\theta\geq0\}}.
\end{align}
This together with (\ref{wp3}) implies
\begin{align}\label{wp5}
\mathbb{E} \|\tilde{\theta}_{k+1}\|^2&\leq \mathbb{E}\|\tilde{\theta}_k\|^2 -\frac{2\beta(1 - p - q)^2}{k} \mathbb{E}| \varphi_k^T \tilde{\theta}_k|+ O(\frac{1}{k^2}).
\end{align}
From Assumption \ref{A2}, we have
\begin{align}\label{wp6}
\|\tilde{\theta}_k\| = \|\hat{\theta}_k - \theta\| \le \|\hat{\theta}_k\| + \|\theta\| \le 2\eta, \quad \forall k \ge 0.
\end{align}
This together with (\ref{12}), by using Cauchy inequality, it is obtained that
\begin{align}\label{wp6}
| \varphi_k^T \tilde{\theta}_k|\leq\| \varphi_k\|\|\tilde{\theta}_k\|\leq2\eta M,
\end{align}
which implies that $\mathbb{E} |\varphi_k^T \tilde{\theta}_k|^2\leq2 \eta M \mathbb{E}| \varphi_k^T \tilde{\theta}_k|$.  Thus, from (\ref{wp5}) it follows that
\begin{align}\label{wp7}
\mathbb{E}\|\tilde{\theta}_{k+1}\|^2 &\le \mathbb{E}\|\tilde{\theta}_k\|^2 - \frac{\beta(1 - p - q)^2\mathbb{E} |\varphi_k^T \tilde{\theta}_k|^2}{\eta M k}+ O\left(\frac{1}{k^2}\right) \cr
&\le \mathbb{E}\|\tilde{\theta}_{k-h+1}\|^2 - \sum_{i=k-h+1}^{k} \frac{\beta(1 - p - q)^2 \mathbb{E}[\tilde{\theta}_i^T \varphi_i \varphi_i^T \tilde{\theta}_i]}{\eta M i}+ \sum_{i=k-h+1}^{k} O\left(\frac{1}{i^2}\right).
\end{align}
\noindent Noticing
\begin{align}\label{wp8}
\tilde{\theta}_i^T \varphi_i \varphi_i^T \tilde{\theta}_i &= \tilde{\theta}_{k-h+1}^T \varphi_i \varphi_i^T \tilde{\theta}_{k-h+1} + ((\tilde{\theta}_i - \tilde{\theta}_{k-h+1})^T \varphi_i)^2  + 2(\tilde{\theta}_i - \tilde{\theta}_{k-h+1})^T \varphi_i \varphi_i^T \tilde{\theta}_{k-h+1} \cr
&\ge \tilde{\theta}_{k-h+1}^T \varphi_i \varphi_i^T \tilde{\theta}_{k-h+1}  + 2(\tilde{\theta}_i - \tilde{\theta}_{k-h+1})^T \varphi_i \varphi_i^T \tilde{\theta}_{k-h+1},
\end{align}
\noindent and from (\ref{wp6}) and Lemma \ref{lemma2}, it follows that,
\begin{align}\label{wp9}
& - \sum_{i=k-h+1}^{k} \frac{\tilde{\theta}_i^T \varphi_i \varphi_i^T \tilde{\theta}_i}{i} \cr
&\le - \sum_{i=k-h+1}^{k} \frac{\tilde{\theta}_{k-h+1}^T \varphi_i \varphi_i^T \tilde{\theta}_{k-h+1}}{i}
+ 4M^2\eta \sum_{i=k-h+1}^{k} \frac{\|\tilde{\theta}_i - \tilde{\theta}_{k-h+1}\|}{i} \cr
&\le - \frac{\delta_{\varphi} h \|\tilde{\theta}_{k-h+1}\|^2}{k} + O\left( \frac{1}{(k-h+1)^2} \right),
\end{align}
\noindent for $k \ge h$. Substituting (\ref{wp9}) into (\ref{wp7}) gives
\begin{align}\label{wp10}
& \mathbb{E}\|\tilde{\theta}_{k+1}\|^2 \cr
\le& \left( 1 - \frac{2\delta_{\varphi} \beta h(1 - p - q)^2}{\eta M k} \right) \mathbb{E}\|\tilde{\theta}_{k-h+1}\|^2 + O\left( \frac{1}{(k-h+1)^2} \right) \cr
\le& \prod_{j=0}^{\lfloor \frac{k}{h} \rfloor - 1} \left( 1 - \frac{2\delta_{\varphi} \beta h(1 - p - q)^2}{\eta M (k-jh)} \right) \mathbb{E}\left\| \tilde{\theta}_{k+1-\lfloor \frac{k}{h} \rfloor h} \right\|^2\cr
&+ \sum_{j=1}^{\lfloor \frac{k}{h} \rfloor} \prod_{i=0}^{j-1} \left( 1 - \frac{2\delta_{\varphi}\beta h(1 - p - q)^2}{\eta M(k-ih)} \right) O\left( \frac{1}{(k-jh+1)^2} \right).
\end{align}
 Let $\kappa = \lceil \frac{k}{h} \rceil - \lfloor \frac{k}{h} \rfloor$. On one hand, the first item on the right side of (\ref{wp10}) is
\begin{align}\label{wp11}
\prod_{j=0}^{\lfloor \frac{k}{h} \rfloor - 1} \left( 1 - \frac{2\delta_{\varphi} \beta h(1 - p - q)^2}{\eta M(k-jh)} \right)
&\le \prod_{m=\kappa+1}^{\lceil \frac{k}{h} \rceil } \left( 1 - \frac{2\delta_{\varphi} \beta h(1 - p - q)^2}{\eta M(mh)} \right) \cr
&= \prod_{m=\kappa+1}^{\lceil \frac{k}{h} \rceil } \left( 1 - \frac{\frac{2\delta_{\varphi} \beta(1 - p - q)^2}{\eta M}}{m} \right) .
\end{align}
\noindent On the other hand, the second item on the right side of (\ref{wp10}) is
\begin{align}\label{wp12}
& \sum_{j=1}^{\lfloor \frac{k}{h} \rfloor } \prod_{i=0}^{j-1} \left( 1 - \frac{2\delta_{\varphi} \beta h (1 - p - q)^2}{\eta M (k-ih)} \right) O\left( \frac{1}{(k-jh+1)^2} \right) \cr
&\le \sum_{j=1}^{\lfloor \frac{k}{h} \rfloor } \prod_{i=0}^{j-1} \left( 1 - \frac{2\delta_{\varphi} \beta h(1 - p - q)^2}{\eta M (k-ih)} \right) O\left( \frac{1}{(\lfloor \frac{k}{h} \rfloor - j + 1)^2 h^2} \right) \cr
&\le  \sum_{m=1}^{\lfloor \frac{k}{h} \rfloor} \prod_{l=\kappa+m}^{\lceil \frac{k}{h} \rceil } \left( 1 - \frac{\frac{2\delta_{\varphi} \beta(1 - p - q)^2}{\eta M }}{l} \right) O\left( \frac{1}{m^2 h^2} \right) \cr
&\le \sum_{m=1}^{\lfloor \frac{k}{h} \rfloor} \prod_{l=\kappa+m}^{\lceil \frac{k}{h} \rceil} \left( 1 - \frac{\frac{2\delta_{\varphi} \beta(1 - p - q)^2}{\eta M }}{l} \right) O\left( \frac{1}{m^2} \right).
\end{align}
\noindent From (\ref{wp10})-(\ref{wp12}) and Lemmas \ref{lemma3}, \ref{lemma4}, when $\frac{2\delta_{\varphi} \beta(1 - p - q)^2}{\eta M } > 1$, we obtain $\mathbb{E}\|\tilde{\theta}_k\|^2 = O(\frac{1}{k})$.

\begin{remark}
Theorem \ref{privacy} shows that the convergence rate is $O(\frac{1}{k})$ when there exists privacy protection, while Theorem \ref{noprivacy} shows that the convergence rate is also $O(\frac{1}{k})$  when there does not exist privacy protection. This indicates that privacy protection is almost free.
\end{remark}
\begin{remark}
In contrast with a fixed threshold of binary quantizer in \cite{guo2013,WangY2021,WangY2026}, by using the time-varying thresholds, Theorem \ref{noprivacy} shows the estimation center can achieve the almost sure convergence and mean-square convergence for the deterministic system (\ref{system}) under data tampering attacks.
\end{remark}
\subsection{Extension to multi-agent systems}
In this section, we consider the following multi-agent systems.
\begin{align}\label{Multisystem}
y_{k,i} = \varphi_{k,i}^{T}\theta,
\end{align}
where $\theta\in\Omega\subseteq\mathbb{R}^{p}$ is the unknown parameter to be estimated, and $\varphi_{k,i}\in \mathbb{R}^{p}$ is the regressor vector of Agent $i$, $i=1,\ldots,n$. To protect $y_{k,i}$, each agent adds the noise based on Theorem 2. Here we consider each agent adds the same variance noise for simplicity. The results still hold for different variance noise of different agents. In this case, the local differential privacy is achieved. For more details about the local differential privacy, please refer to Subsection F of main results in \cite{WangJM2024}. Let $\omega_{k,i}$ be a privacy noise sequence. The private output $\tilde{y}_{k,i}=y_{k,i}+\omega_{k,i}$ is measured by a binary-valued sensor with the time-varying threshold of $\varphi_{k,i}^{T}\hat{\theta}_{k,i}$, where $\hat{\theta}_{k,i}$ is the estimate of $\theta$, and is represented by the following indicator function:
\begin{equation}\label{msk}
s_{k,i}^0 = I_{\{\tilde{y}_{k,i}\leq \varphi_{k,i}^{T}\hat{\theta}_{k,i}\}}=\left\{\begin{array}{rl}
                            1,& \tilde{y}_{k,i}\leq \varphi_{k,i}^{T}\hat{\theta}_{k,i};\\
                            0,& \mbox{otherwise}.
                          \end{array}\right.
\end{equation}
$s_{k,i}^0$ is transmitted to the remote estimation center subject to data tampering attacks. The outputs when attacker exists, denoted as $s_{k,i}$. The data tampering attacker model considered here is defined as
\begin{equation}\label{mog}
   \left\{
   \begin{array}{lll}
     \mathbb{P}\{s_{k,i}=0|s_{k,i}^0=1\} =p; \\[.1cm]
     \mathbb{P}\{s_{k,i}=1|s_{k,i}^0=1\} =1-p; \\[.1cm]
     \mathbb{P}\{s_{k,i}=1|s_{k,i}^0=0\} =q; \\[.1cm]
     \mathbb{P}\{s_{k,i}=0|s_{k,i}^0=0\} =1-q.
   \end{array}
   \right.
\end{equation}

In this section, we aim to propose a distributed recursive projection algorithm such that the differential privacy, data tampering attacks, and quantized measurements are considered in a unified framework. The detailed steps are given in Algorithm 2.
\begin{algorithm}[H]
\caption{A privacy preserving distributed recursive projection algorithm}
\begin{algorithmic}
\STATE Initial estimate $\hat{\theta}_{1,i} \in \Omega$, the step-size sequence $\{b_k\}_{k \geq 1}$, a constant gain parameter $\beta>0$\\

\STATE For $k = 1, 2, \ldots$\\

\STATE Compute
\begin{align}\label{M10}
\tilde{s}_{k,i} &= \beta(1 - (p + q)) \left( (1 - (p + q))F(0) + q - s_{k,i} \right).
\end{align}
\STATE Update parameter estimate,
\begin{align}\label{M11}
\hat{\theta}_{k+1,i} = \Pi_{\Omega} \left\{ \hat{\theta}_{k,i} + b_k \sum_{j\in\mathcal{N}_{i}}a_{ij}(\hat{\theta}_{k,j}-\hat{\theta}_{k,i})+ b_k \varphi_{k,i} \tilde{s}_{k,i} \right\}.
\end{align}
\end{algorithmic}
\end{algorithm}

To show the convergence of Algorithm 2, the following assumptions are needed.
\begin{assumption}\label{M1}
The graph is undirected and connected.
\end{assumption}
\begin{assumption}\label{M3}
(a) The regressor sequence $\{ \varphi_{k,i} \}$ satisfies $\| \varphi_{k,i} \| \leq M < \infty$.
(b) There exist an integer $h \geq p$ and a constant $\delta > 0$ such that
\begin{align}\label{M13}
\mathbb{E} \left[\sum_{i=1}^{n} \sum_{l=k}^{k+h} \varphi_{l,i} \varphi_{l,i}^T \bigg| \mathcal{\tilde{F}}_{k-1} \right] \geq \delta I, \quad \forall k \geq 1,
\end{align}
where $I$ is the $p \times p$ identity matrix and the natural filtration $\mathcal{\tilde{F}}_{k}$ is defined as
\begin{align*}
\mathcal{\tilde{F}}_{k}=\sigma\{\varphi_{0,i}, \ldots, \varphi_{k+1,i}, \omega_{0,i}, \ldots, \omega_{k,i}, i=1,\ldots,n\}, k\geq0.
\end{align*}.
\end{assumption}
\begin{assumption}\label{M4}
The noise $ \omega_{k,i} $ is independent of $\varphi_{0,i}, \ldots, \varphi_{k,i}, \omega_{0,i}, \ldots, \omega_{k-1,i}$, and the density function $f_{i}(\cdot)$ of the noise $ \omega_{k,i} $ satisfies
\begin{align}
\underline{f} := \min_{1\leq i\leq n} \inf_{|x| \leq M\eta} f_{i}(x) > 0.
\end{align}
\end{assumption}

Denote $\tilde{\theta}_{k,i}=\hat{\theta}_{k,i}-\theta$, $\hat{\Theta}_{k}=(\hat{\theta}_{k,1}^{T},\ldots,\hat{\theta}_{k,n}^{T})^{T}$, $\tilde{\Theta}_{k}=(\tilde{\theta}_{k,1}^{T},\ldots,\tilde{\theta}_{k,n}^{T})^{T}$, $\Phi_k={\rm diag}\{\varphi_{k,1},\ldots,\varphi_{k,n}\}$, $\tilde{s}_{k}=(\tilde{s}_{k,1}^T,\ldots,\tilde{s}_{k,n}^T)^T$,
${\bf \Pi}\{\cdot\}$ as ${\bf \Pi}\{\zeta\}=(\Pi\{\zeta_{1}\},\ldots,\Pi\{\zeta_{n}\})^{T}$ for $\zeta=(\zeta_1^{T},\ldots,\zeta_n^T)^T$.
Then, (\ref{M11}) is rewritten as the following compact form.
\begin{align}\label{M14}
\hat{\Theta}_{k+1} = {\bf \Pi}\left\{(I-b_k\mathcal{L}\otimes I_p) \hat{\Theta}_{k}+ b_k \Phi_{k} \tilde{s}_{k}\otimes 1_p \right\}.
\end{align}
\begin{theorem}\label{thml}
Under Assumptions \ref{A2}, \ref{A5}-\ref{M4}, if $p+q\neq1$, then the parameter estimate generated by Algorithm 2 satisfies
\begin{align}
\lim_{k \to \infty} \mathbb{E} \| \tilde{\theta}_{k,i} \|^2 = 0.
\end{align}
Moreover, if $\sum_{k=1}^{\infty} b_k^2 < \infty$, then the estimate $\hat{\theta}_{k,i}$ also converges almost surely to the true parameter, i.e.,
\begin{align}
\lim_{k \to \infty} \tilde{\theta}_{k,i} = 0, \quad \text{a.s.}
\end{align}
\end{theorem}
{\bf Proof.}
By $\tilde{s}_{k,i}^2 \leq \beta^2$, (\ref{pr1}), (\ref{M10}), and (\ref{M14}), we have
\begin{align}\label{mL1}
\|\tilde{\Theta}_{k+1}\|^2 &\leq \|(I-b_k\mathcal{L}\otimes I_p) \tilde{\Theta}_{k}+ b_k \Phi_{k} \tilde{s}_{k}\otimes 1_p\|^2 \cr
&\leq (1-b_k\lambda_2({\mathcal{L}}))\|\tilde{\Theta}_k\|^2 + 2b_k (\tilde{s}_{k}\otimes 1_p)^T \Phi_k^T (I-b_k\mathcal{L}\otimes I_p)\tilde{\Theta}_k + nb_k^2 \|\Phi_k\|^2 \beta^2.
\end{align}
Similar to (21), it is obtained that
\begin{align}\label{ML2}
\mathbb{E}\left[2b_k (\tilde{s}_{k}\otimes 1_p)^T \Phi_k^T  (I-b_k \mathcal{L}\otimes I_p)\tilde{\Theta}_k|\mathcal{\tilde{F}}_{k-1}\right]&\leq -2b_k \beta(1-p-q)^2\underline{f}\tilde{\Theta}_k^T\Phi_k\Phi_k^T  (I-b_k \mathcal{L}\otimes I_p)\tilde{\Theta}_k.
\end{align}
This together with (\ref{mL1}) implies
\begin{align}\label{ML3}
&\mathbb{E}\left[\|\tilde{\Theta}_{k+1}\|^2|\mathcal{\tilde{F}}_{k-1}\right]\cr
&\leq (1-b_k\lambda_2({\mathcal{L}}))\|\tilde{\Theta}_k\|^2 + \mathbb{E}\left[2b_k (\tilde{s}_{k}\otimes 1_p)^T \Phi_k^T (I-b_k \mathcal{L}\otimes I_p)\tilde{\Theta}_k|\mathcal{\tilde{F}}_{k-1}\right] + nb_k^2 \|\Phi_k\|^2 \beta^2\cr
&\leq (1-b_k\lambda_2(\mathcal{L}))\|\tilde{\Theta}_k\|^2-2b_k \beta(1-p-q)^2\underline{f}\mathbb{E}\left[\tilde{\Theta}_k^T\Phi_k\Phi_k^T (I-b_k \mathcal{L}\otimes I_p)\tilde{\Theta}_k|\mathcal{\tilde{F}}_{k-1}\right] + nb_k^2 \|\Phi_k\|^2 \beta^2\cr
&\leq (1-b_k\lambda_2(\mathcal{L}))\|\tilde{\Theta}_k\|^2-2b_k \beta(1-p-q)^2\underline{f}\tilde{\Theta}_k^T\Phi_k\Phi_k^T\tilde{\Theta}_k + O\left(b_k^2 \right)
\end{align}
Similar to Lemma 2, we have $\|\tilde{\theta}_{k+l,i}-\tilde{\theta}_{k,i}\|=O(b_{k+l}), \forall k, l \in \mathbb{N}$. Based on this, from (\ref{ML3}) it follows that
\begin{align}\label{ML5}
&\mathbb{E}\|\tilde{\Theta}_{k+1}\|^2\cr
\leq& \mathbb{E}\left[(1-b_k\lambda_2(\mathcal{L}))\|\tilde{\Theta}_k\|^2-2b_k \beta(1-p-q)^2\underline{f}\tilde{\Theta}_k^T\Phi_k\Phi_k^T \tilde{\Theta}_k\right]+ O\left(b_k^2\right)\cr
\leq& (1-b_k\lambda_2(\mathcal{L}))\mathbb{E}\|\tilde{\Theta}_{k-h}\|^2-2b_k \beta(1-p-q)^2\underline{f}\mathbb{E}\left[\tilde{\Theta}_{k-h}^T\sum_{l=k-h}^{k}\Phi_l\Phi_l^T \tilde{\Theta}_{k-h}\right] + O\left(b_k^2\right)\cr
\leq& \left((1-b_k\lambda_2(\mathcal{L}))^{2}-2b_k \beta(1-p-q)^2\underline{f}\delta\right)\mathbb{E}\|\tilde{\Theta}_{k-h}\|^2+  O\left(b_k^2\right).
\end{align}
This further implies that
\begin{align}\label{ML6}
\mathbb{E}\|\tilde{\Theta}_{k+1}\|^2
\leq& \left((1-b_k\lambda_2(\mathcal{L}))^{2}-2b_k \beta(1-p-q)^2\underline{f}\delta\right)\mathbb{E}\|\tilde{\Theta}_{k-h}\|^2+  O\left(b_k^2\right).
\end{align}
Then, based on Lemma \ref{lemma1} and Assumption \ref{A5}, and noting  $\sum_{k=1}^{\infty} b_k = \infty$ and $\lim_{k \to \infty} b_k= 0$, it follows that $\lim_{k\rightarrow\infty} \mathbb{E}\left[\|\tilde{\Theta}_{k}\|^2\right]=0$.

On the other hand, by (\ref{ML3}) we have $\mathbb{E}[\|\tilde{\Theta}_{k+1}\|^2 | \mathcal{\tilde{F}}_{k-1}] \leq \|\tilde{\Theta}_k\|^2 +O(b_k^2)$, which together with Lemma 1.2.2 in \cite{chen2002} and $\sum_{k=1}^{\infty} b_k^2 < \infty$ implies that $\|\tilde{\Theta}_k\|$ converges to a bounded limit a.s. Note that $\lim_{k \to \infty} \mathbb{E}\|\tilde{\Theta}_k\|^2 = 0$. Then,  by Theorem 5.2.1 of \cite{Gut2005}, $\tilde{\Theta}_k$ almost surely converges to $0$. \hfill$\square$
\begin{remark}
In Theorem \ref{thml}, the mean square and almost sure convergence is given for multi-agent systems with quantized measurements, differential privacy-preserving, and data tampering attacks. Quantized measurements are considered in this paper while quantized communications (e.g. \cite{WangY2026}) are another interesting works which worthy to be studied in the future.
\end{remark}
\section{Numerical Simulation}\label{sec:5}

Consider the system $y_{k} = \varphi_k^T\theta$ with the binary observation  $s_{k}^0 = I_{\{\tilde{y}_k\leq \varphi_k^T\hat{\theta}_k\}}$, under the data  tampering attacks \eqref{og}, where $\theta=[3,-1]^{T}$ is to be estimated and is known as in $\Omega=\{(x,y):|x|<6,|y|<6\}$. The privacy
noise $\omega_{k}$ follows the Gaussian distribution with $\epsilon=0.2$ and $\Delta_k=0.2$. The inputs $\varphi_k=\{u_k,u_{k-1}\}$ with $u_k$ following the uniform distribution of $N(0, 2)$. Algorithm 1 has a step size of $\beta=100$, $b_k=1/k$, and an initial value of $\theta_1=[1,1]^{T}$. All the simulations are looped 50 times. Figure 2 shows the estimation error of the estimation center as $\epsilon=0.2, p = 0.2, q = 0.3$ and $\epsilon=0.2, p = 0.8, q = 0.9$, respectively. From Figure 2, it can be seen that even when the data tampering attacks \eqref{og} is close to 1, the estimation center's estimation error still converges to 0.
\begin{figure}[H]
\centering
\includegraphics[width=0.7\linewidth]{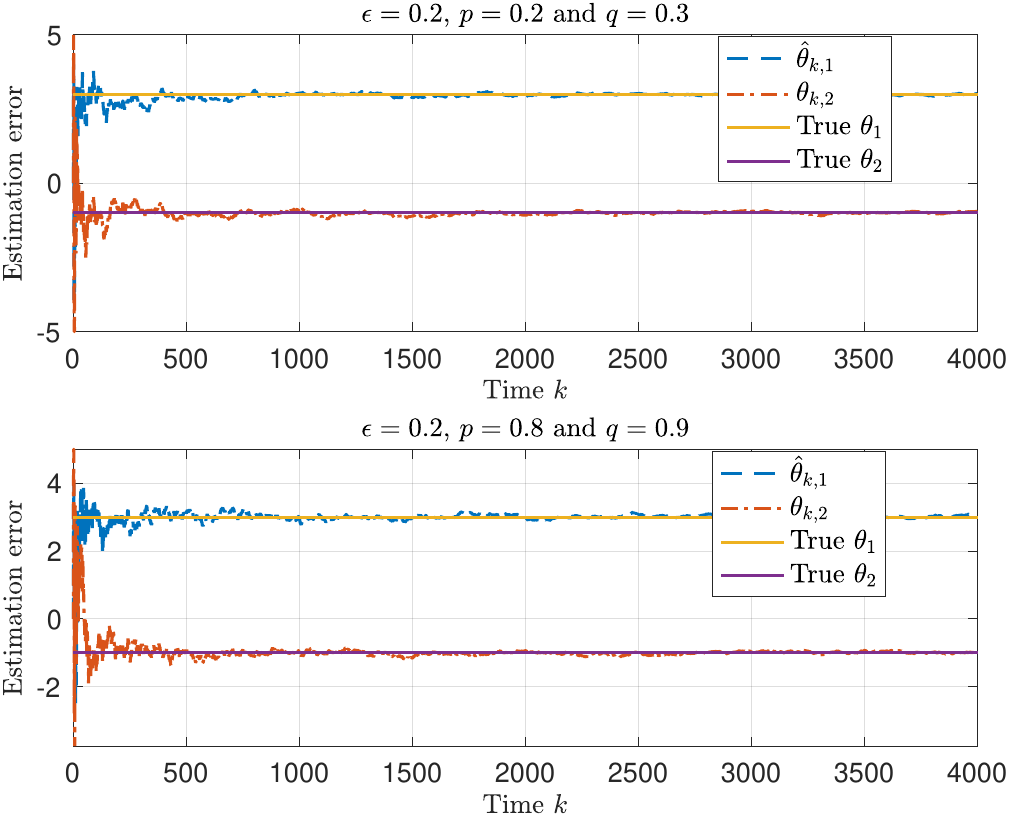}
\caption{The estimation center's estimation error}\label{fig1}
\end{figure}
\subsection{Multi-agent systems}
Consider the 5-agent systems $y_{k,i} = \varphi_{k,i}^{T}\theta$ with the binary observation  $s_{k,i}^0 = I_{\{\tilde{y}_{k,i}\leq \varphi_{k,i}^T\hat{\theta}_{k,i}\}}$, $i=1,2,\ldots,5$, under the data tampering attacks \eqref{mog}.
The communication graph is given in Figure 3, and the adjacent matrix is given as follows.
\begin{equation*}
	 	A = \begin{bmatrix}
	 		0.0 & 0.5 & 0.0 & 0.0 & 0.5 \\
	 		0.5 & 0.0 & 0.5 & 0.0 & 0.0 \\
	 		0.0 & 0.5 & 0.0 & 0.5 & 0.0 \\
	 		0.0 & 0.0 & 0.5 & 0.0 & 0.5 \\
	 		0.5 & 0.0 & 0.0 & 0.5 & 0.0
	 	\end{bmatrix}
\end{equation*}
\begin{figure}[H] 
\centering
\includegraphics[width=0.26\linewidth]{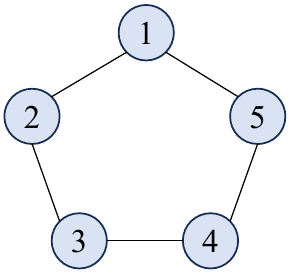} 
\caption{Communication graph} 
\label{fig:example}
\end{figure}
 $\theta=[3,-1]^{T}$ is to be estimated and is known as in $\Omega=\{(x,y):|x|<6,|y|<6\}$. The privacy noise $\omega_{k}$ follows the Gaussian distribution with $\epsilon=1$ and $\Delta_k=1$. The inputs $\varphi_{k,i}=\{u_{k,i},u_{k-1,i}\}$ with $u_{k,i}$ following the uniform distribution of $N(0, 2)$. Algorithm 2 has a step size of $\beta=100$, $b_k=1/k$, and an initial value of $\theta_{1,i}=[1,1]^{T}$. All the simulations are looped 50 times. Figure 4 shows the estimation error of the estimation center as $p = 0.2, q = 0.4$ and $p = 0.7, q = 0.9$, respectively. From Figure 4, it can be seen that even when the tampering attacks \eqref{mog} is close to 1, the estimation center's estimation error still converges to 0.
\begin{figure}[H] 
\centering
\includegraphics[width=0.8\linewidth]{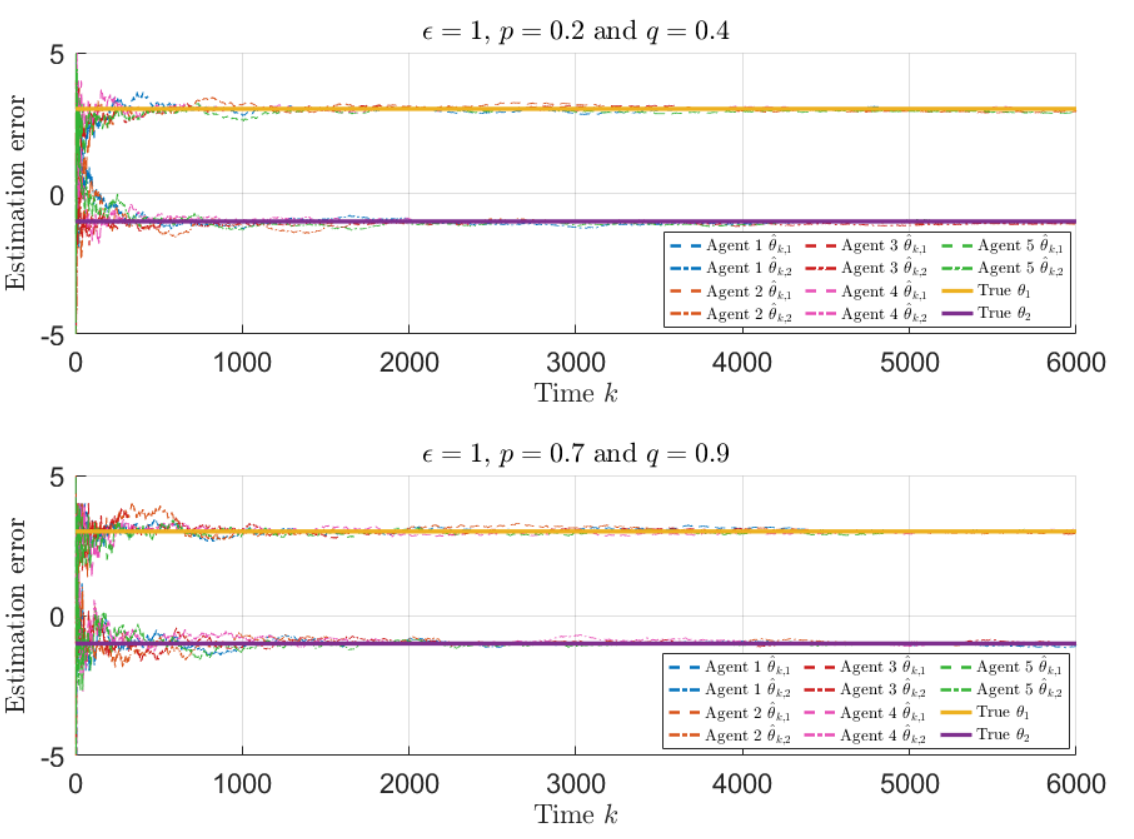} 
\caption{Each agent's estimation error} 
\label{fig:example}
\end{figure}

\section{Conclusions}\label{sec:6}
This paper considers the system identification under communication constraints, privacy requirements and data tampering attacks. Regarding the communication costs, the sensor transmits only 1 bit of information to the remote estimation center at each time step. Regarding the privacy protection, differential privacy is used, and the effect of the differential privacy is shown by analyzing the convergence rate of the algorithm. A recursive projection algorithm is proposed such that the estimation center's estimation error converges to 0 in the almost sure and mean-square sense. Then, the problem is further extended to the multi-agent systems, and a distributed recursive projection algorithm is proposed such that each agent estimates the unknown parameter in the almost sure and mean-square sense.

Note that deception attacks \cite{LiXM2025} and DoS attacks \cite{SuW2025} are  two other common attack strategies. Then, how to extend the method in this paper to defense these attack strategies is worthy to be studied in the further.

\subsection*{Acknowledgements}

The work was supported by National Natural Science Foundation of China under Grants No. 62433020, No. 62573044, and No. 62588101, and CAS Project for Young Scientists in Basic Research, China under Grant YSBR-008.

\end{document}